\documentclass[aps,prb,twocolumn]{revtex4-2}
\usepackage{amssymb} 
\usepackage{amsmath}
\usepackage{multirow}
\usepackage{booktabs}
\usepackage{tabularx} 
\usepackage{graphicx}% Include figure files
\usepackage{dcolumn}% Align table columns on decimal point
\usepackage{bm}% bold math
\usepackage{makecell}
\usepackage[version=4]{mhchem}
\begin{document}
	
	\preprint{APS/123-QED}
	
	\title{\textbf{Bidirectional Optimization onto Thermoelectric Performance via Hydrostatic-Pressure in Chalcopyrite AgXTe$_2$ (X=In, Ga)}}% Force line breaks with \\
    \thanks{Corresponding author: liuyanhui@nub.edu.cn,cuitian@nub.edu.cn}%
	
    \author{Siqi Guo,$^{1}$ Jincheng Yue,$^{1}$ Jiongzhi Zheng,$^{2,3}$ Hui Zhang,$^{1}$, Ning Wang,$^{1}$, Junda Li,$^1$ Yanhui Liu,$^{1\ast}$ and Tian Cui,$^{1,4\ast}$}
    \affiliation{$^{1}$Institute of High-Pressure Physics, School of Physical Science and Technology, Ningbo University, Ningbo 315211, China} 
    \affiliation{$^{2}$Thayer School of Engineering, Dartmouth College, Hanover, New Hampshire, 03755, USA}
    \affiliation{$^{3}$Department of Mechanical and Aerospace Engineering, The Hong Kong University of Science and Technology, Clear Water Bay, Kowloon, 999077, Hong Kong} 
    \affiliation{$^{4}$State Key Laboratory of Superhard Materials, College of Physics, Jilin University, Changchun 130012, China} 
	
	\date{\today}% It is always \today, today,

\begin{abstract}
  The pursuit of high-performance thermoelectric materials has driven extensive research into diverse material systems and optimization approaches. Pressure tuning has emerged as a powerful strategy for manipulating the thermoelectric properties of materials by inducing structural and electronic modifications. Herein, we systematically investigate the transport properties and thermoelectric performance concerning lattice distortions induced by hydrostatic pressure in Ag-based chalcopyrite AgXTe$_2$ (X=In, Ga). The findings reveal that the lattice distortion in AgXTe\textsubscript{2} exhibits distinct behaviors under lattice compression, diverging from trends observed at ambient pressure. Importantly, the hydrostatic pressure breaks the phenomenally negative correlation between thermal conductivity and lattice distortion. Pressure-induced softening of low-frequency acoustic phonons broadens the low-energy phonon spectrum, enhancing interactions between acoustic and optical phonons. Such broadening substantially increases the number of available three-phonon scattering channels, resulting in a marked reduction in thermal conductivity. Meanwhile, we establish a macroscopic connection between metavalent bonding and anharmonicity, providing an indirect explanation for lattice anharmonicity through pressure-driven transferred charge. Additionally, the applied pressure achieves a notable net increase in the power factor despite the strong coupling of electrical transport parameters, which underscores the potential for bidirectional optimization of transport properties in AgXTe$_2$. As a result, the maximum \textit{ZT} value of AgInTe$_2$ is nearly doubled, demonstrating that pressure modulation is a powerful strategy for enhancing thermoelectric performance. Our work not only establishes the link between pressure, lattice dynamics, and thermoelectric properties within chalcopyrite AgXTe$_2$, but also inspires the exploration of pressure-related optimization strategies for conventional thermoelectric materials.

\end{abstract}
	
	\maketitle
	
	%\tableofcontents
	
	\section{\label{sec:level1}INTRODUCTION}
 \par Thermoelectric (TE) materials belong to the distinctive category of functional materials capable of interconverting thermal and electrical energies~\cite{qijian,te-handbook,yue2024o-CsCu5S3,azimi2014towards}. The dimensionless figure of merit \textit{ZT}=\textit{S}$^{2}$$\sigma$\textit{T}/($\kappa_{e}$+$\kappa_{L}$), where the \textit{S}, $\sigma$, $\kappa_{e}$, and $\kappa_{L}$ represent the Seebeck coefficient, electrical conductivity,  electronic and lattice thermal conductivity, respectively, is used to evaluate thermoelectric properties~\cite{te-handbook,Complex,ZhangCoP3}. An ideal thermoelectric material should simultaneously exhibit several key characteristics: high electrical conductivity to minimize internal energy dissipation, a substantial Seebeck coefficient to generate significant voltage, and low thermal conductivity to maintain a robust temperature gradient. The intricate coupling among these parameters, however, poses a formidable challenge to the optimization of $ZT$~\cite{chang2018anharmoncity,shi2016recent,gaoAgSbSe22018extraordinary}. 
 
 \par A promising strategy to address this challenge is the application of hydrostatic pressure~\cite{li2024high,deng2024high}. The application of external pressure can induce significant changes in the crystal and electronic structures of materials, which may lead to the decoupling of the fundamental physical parameters, including $S$, $\sigma$, $\kappa_e$, and $\kappa_L$, thereby enabling the simultaneous optimization of electronic and lattice transport properties, a phenomenon that has emerged as a critical factor in modulating the thermoelectric material properties, as recent scientific advancements have evidenced~\cite{zhou2022pressurethermal,shi2019unprecedented}. Experimental investigations have revealed that in the pressure range up to 30 GPa, the thermal conductivity of BAs shows an anomalous trend of increasing and then decreasing with pressure, and this complex behavior is attributed to the competing relationship between the three-phonon and four-phonon processes dominating the scattering under the influence of pressure~\cite{BAs2022anomalous}. Moreover, studies have demonstrated that the application of pressure to SnSe induces the formation of additional conduction band valleys, thereby enhancing band degeneracy and consequently improving its thermoelectric properties~\cite{SnSepressure,zhang2016pressure}. These findings demonstrate the potential of pressure in enhancing the thermoelectric performance of materials by modifying their crystallographic and electronic configurations.
	
 \par Diamond-like materials with low-symmetry structures have garnered significant interest in the fields of electronics and optoelectronics, owing to their broad electronic band gaps~\cite{opticalais,ferhati2020optics,stock2020pure,ishizuka2024aluminum}. It is noteworthy that these materials also exhibit remarkable thermoelectric properties, as exemplified by ternary chalcopyrite~\cite{AgInSe22018intrinsically,li2021anomalous,plirdpring2012chalcopyrite}. Experimental results have demonstrated that the incorporation of Ag and Ga into the Cu$_{1-x}$Ag$_x$InTe$_2$ and (Cu$_{1-x}$Ag$_x$)(In$_{1-y}$Ga$_y$)Te$_2$ systems have resulted in a maximum $\textit{ZT}$ value of 1.6, representing the highest value reported to date~\cite{xie2020ultralow,xie2021ultralow,zhang2019design}. This exceptional performance has been attributed to the unique electronic property of chalcopyrite compounds, which exhibit full conformity to the recently defined metavalent bonding region, characterized by pronounced anharmonicity~\cite{2021metavalent}. Further investigation revealed that the high performance observed in chalcopyrite compounds AgGaTe$_2$ was attributed to strong phonon scattering caused by distorted Ag tetrahedra, a consequence of weak $sd^3$ orbital hybridization in Ag atoms~\cite{xie2022hidden}. These insights have deepened our understanding of chalcopyrite compounds and have led to a growing interest in the structure of diamondoid chalcopyrites. 
 At present, anomalous thermal transport of chalcopyrite structures at high pressures has been reported. ~\cite{li2021anomalous,yu2020combined,yue2023pressure,yu2018large}For example, a negative correlation between pressure and thermal conductivity has been observed in CuInTe$_2$~\cite{yue2023pressure,yu2018large}. While these findings provide valuable insights, the microscopic mechanisms underlying the anomalous thermal transport behavior under pressure remain inadequately understood, and there is a need for further comprehensive and systematic studies to uncover the underlying physical principles.	
 \par In this study, we comprehensively investigated the thermal/electron transport and thermoelectric properties under the lattice distortions caused by hydrostatic pressure in Ag-based chalcopyrites AgXTe$_2$ (X=In, Ga). This study reveals lattice distortion in AgXTe$_2$ at ambient pressure, suggesting a high phonon scattering rate due to the structural complexity. Under applied pressure, the distortion in AgInTe$_2$ remains relatively constant, while that of AgGaTe$_2$ increases monotonically. We calculated the lattice thermal conductivity of both compounds under various pressures and found that the correlation between lattice distortion and thermal conductivity is decoupled at pressures. Further analysis reveals a significant increase in the three-phonon scattering phase space for AgInTe$_2$ under pressure, which is identified as the primary driver of the substantial change in $\kappa_{L}$. Moreover, the impact of pressure on electronic transport properties was investigated, demonstrating a considerable enhancement of carrier mobility considering three dominant scattering mechanisms. Notably, a positive correlation between charge transfer and anharmonicity is observed within the compounds, with pressure-induced charge transfer leading to a substantial increase in anharmonicity. These findings suggest that the application of pressure can effectively enhance the overall thermoelectric performance of chalcopyrite AgXTe$_2$ (X=In, Ga) by simultaneously optimizing both thermal and electronic transport properties.

	\begin{table*}[t]
		\centering
		\caption{Structural parameters of \ce{AgInTe2} and \ce{AgGaTe2} under different pressures: lattice constants (a/b, c), bond lengths (Ag-Te, In/Ga-Te), tetragonal distortion ($\eta$).}
		\label{tbl:1}
		\setlength{\tabcolsep}{5mm}{
			\begin{tabular}{lccccccc}
				\hline
				Compound & Pressure (GPa) & a/b (\AA) & c (\AA) & Ag-Te (\AA) & In/Ga-Te (\AA) & $\eta$ \\
				\hline
				\ce{AgInTe2} & 0 & 6.57 & 13.02 & 2.81 & 2.86 & 0.019  \\
				& 4 & 6.39 & 12.64 & 2.72 & 2.79 & 0.020  \\
				\ce{AgGaTe2} & 0 & 6.41 & 12.34 & 2.80 & 2.68 & 0.073  \\
				& 3.8 & 6.26 & 11.99 & 2.73 & 2.62 & 0.083  \\
				\hline
		\end{tabular}}
	\end{table*}
	\begin{figure*}[t]
		\centering
		\includegraphics[width=0.95\textwidth]{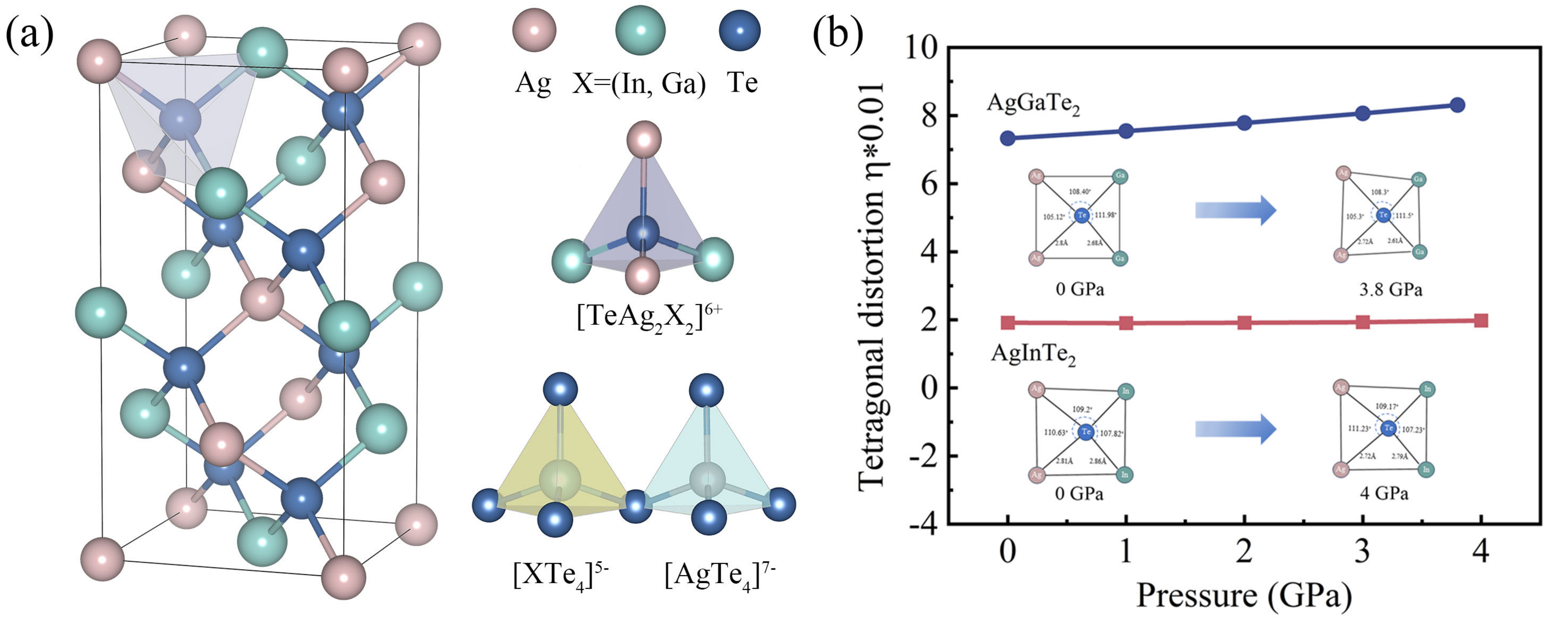}
		\caption{(a)  Schematic representation of the chalcopyrite crystal structure. Adjacent to the structure are three smaller tetrahedra, representing the coordination environments of [\ce{TeAg2X2}]$^{6+}$, [\ce{XTe4}]$^{5-}$, and [\ce{AgTe4}]$^{7-}$, respectively. These tetrahedral configurations illustrate the local bonding arrangements within the crystal structure.(b) Pressure dependence of the tetragonal distortion parameter $\eta$ for the chalcopyrite compounds \ce{AgInTe2} and \ce{AgGaTe2}. The inset illustrates the pressure-induced distortion of the [\ce{TeAg2X2}]$^{6+}$ tetrahedral unit in \ce{AgXTe2}.}
		\label{Fig:1}
	\end{figure*}
	
	\section{COMPUTATIONAL METHODS} 
    \subsection{Density-function theory calculation}
 \par Based on the framework of density functional theory (DFT), first-principles calculations are performed in the VASP software package using the projector augmented wave (PAW) method~\cite{kresse1996efficient,blochl1994projector,hohenberg1964inhomogeneous}. We used the generalized gradient approximation (GGA) of the Perdew-Burke-Ernzerhof (PBE) functional to describe the exchange-correlation interaction~\cite{1996generalized,GGA1996}. A plane-wave cut-off of 600 eV was used in all the calculations. We considered the AgXTe$_2$ (X=In, Ga) compound with a chalcopyrite-type structure, which has a unit cell containing 4 Ag atoms, 4 In (or Ga) atoms, and 8 Te atoms. The project augmented wave pseudopotentials were used to treat the Ag (4\textit{d$^{10}$}5\textit{s$^1$}), In (5\textit{s$^2$}5\textit{p$^1$}), Ga (4\textit{s$^2$}4\textit{p$^1$}) and Te (5\textit{s$^2$}5\textit{p$^4$}) shells as valence states. The geometry optimization and phonon calculations were performed with the total energy and forces convergence criteria of 10$^{-8}$ eV and 10$^{-6}$ eV/\AA\, respectively. To overcome the severe band gap underestimation of the PBE functional for semiconductor materials, we employed the HSE06 hybrid functional to investigate the band gap and electronic band structure of AgGaTe$_2$~\cite{HSE06,yue-1TAu}. The static dielectric constant of AgGaTe$_2$ was obtained from the hybrid functional calculations at a dense 8$\times$8$\times$8 \textit{k}-point mesh.

 \subsection{Interatomic force constants and thermal transport}
\par The relaxed crystal structure was then used in the phonon calculations, which were carried out by the Phonopy software package~\cite{togo2015phonon}. The second-order force constant and the phonon dispersion were obtained using a 2$\times$2$\times$1 supercell with 6$\times$6$\times$6 \textit{k}-grid. To extract the third-order interatomic force constant, we used the Thirdorder software package and generated 780 displacement cells by considering 9 adjacent atoms~\cite{PhysRevB.86.174307}. The lattice thermal conductivity is calculated by iteratively solving the phonon Boltzmann transport equation as implemented in the ShengBTE package with the 12$\times$12$\times$12 \textit{q}-grid~\cite{2014shengbte}. The $\kappa_{L}$ tensor can be expressed as	\begin{equation}
	\kappa _{L}^{\alpha \beta } \left (  T\right ) =\frac{1}{N\Omega }  {\textstyle \sum_{q,j}}C_{q,j}\left ( T \right )\upsilon _{q,j}^{\alpha }\upsilon _{q,j }^{\beta } \tau _{q,t} \left (  T\right )
	\end{equation}
where $N$, $\Omega$, $C_{q,j}(T)$, and $\upsilon_{q,j}$ represent the total number of sampling, the volume system, specific heat capacity, and group velocity, respectively. The $\tau_{q,t}$ is defined as the inverse of the scattering rate. Specifically, the three-phonon scattering probability $\Gamma _{\lambda \lambda ^{'} \lambda ^{''} }^{+}$ can be expressed as,
	\begin{equation}
		\begin{aligned}
			\Gamma _{\lambda \lambda' \lambda'' }^{\pm} &= \frac{\hbar \pi }{8N} \left\{
			\begin{matrix}
				2(f_{\lambda'} - f_{\lambda''}) \\
				f_{\lambda'} + f_{\lambda''} + 1 
			\end{matrix} 
			\right\} \\
			&\quad \times \frac{\delta (\omega_{\lambda} \pm \omega_{\lambda'} - \omega_{\lambda''})}{\omega_{\lambda} \omega_{\lambda'} \omega_{\lambda''}} \left| V_{\lambda \lambda' \lambda'' }^{\pm} \right|^{2}
		\end{aligned}
	\end{equation}
	where the upper (lower) row in curly brackets goes with the + (-) sign for absorption (emission) processes, respectively. The scattering matrix elements V$_{\lambda \lambda ^{'} \lambda ^{''}  }^{\pm}$ depend on the third order interatomic force constants (IFCs).
    
    \subsection{Electronic properties and electron-phonon coupling}
    \par To evaluate the electrical transport properties, we utilized the AMSET code~\cite{Amset2021efficient}, which allowed for accurate calculations of electron scattering processes and relaxation times. The differential scattering rate for transitions from an initial electronic state $\Psi_{nk}$ to final states $\Psi_{m\mathbf{k+q}}$, accounting for both inelastic and elastic scattering events, is computed using the following expressions. For inelastic scattering processes, such as polar optical phonon (POP) interactions, the scattering rate is given by:
	\begin{equation}
		\begin{aligned}
			\tau_{nk \to mk+q}^{-1} &= \frac{2\pi}{\hbar} \left| g_{nm}(k,q) \right|^2 \\
			&\quad \times \left[ (n_{po} + 1 - f_{mk+q}) \delta(\varepsilon_{nk} - \varepsilon_{mk+q} - \hbar \omega_{po}) \right. \\
			&\quad \left. + (n_{po} + f_{mk+q}) \delta(\varepsilon_{nk} - \varepsilon_{mk+q} + \hbar \omega_{po}) \right]
		\end{aligned}
	\end{equation}
   Here, $n_{po}$ denotes the phonon occupation number, $f_{mk+q}$ represents the Fermi-Dirac distribution function, and $\omega_{po}$ is the polar optical phonon frequency. The delta functions enforce energy conservation for phonon absorption and emission processes.
   The $\varepsilon_{nk}$ and $\varepsilon_{mk+q}$ represent the energies of the initial and final electronic states, respectively. 
   \par For elastic scattering, such as acoustic deformation potential (ADP) and ionized impurity (IMP) scattering, the differential scattering rate is described by \begin{equation}
		\tau _{nk\to mk+q}^{-1} = \frac{2\pi}{\hbar} \left| g_{nm}(k,q) \right|^{2} 
		\delta \left( \varepsilon_{nk} - \varepsilon_{mk+q} \right)
      \end{equation}
    In this case, the delta function ensures energy conservation without phonon exchange.
	
   \par The electron-phonon coupling matrix element $g_{nm}(\textit{\textbf{k}},\textit{\textbf{q}})$ was computed using a densely interpolated grid of 95$\times$95$\times$49 \textit{k}-points, combined with material parameters obtained from first-principles calculations. This matrix element incorporates contributions from three primary scattering mechanisms: ADP, POP, and IMP. 
	\begin{equation}
		g^{ADP}(k,q)=[\frac{k_{B}T\varepsilon^{2}_{d}}{B} ]^{(\frac{1}{2})} \left \langle \psi _{nk} | \psi _{mk+q} \right \rangle 
	\end{equation}
 
	\begin{equation}
		g^{IMP}(k,q)=[\frac{e^{2}n_{ii}}{\varepsilon _{s}^{2}}] \frac{\left \langle \psi_{nk}  | \psi_{mk+q} \right \rangle }{\left |q\right|^{2} +\beta ^{2}} 
	\end{equation}	
 
	\begin{equation}
		g^{POP}(k,q)=\left [ \frac{\hbar \omega _{po}}{2}(\frac{1}{\varepsilon _{\infty} }-\frac{1}{\varepsilon _{s} }) \right ] \frac{\left \langle \psi_{nk} |\psi _{mk+q} \right \rangle }{\left | q \right | }  
	\end{equation}	
 where $k_B$, $\varepsilon_d$, $B$, $n_{ii}$, $\beta$, $\omega_{po}$, $\varepsilon_{\infty}$ and $\varepsilon_s$ represent the Boltzmann constant, deformation potential, bulk modulus, concentration of ionized impurities, inverse screening length, optical phonon frequency, high-frequency dielectric constant, and static dielectric constant, respectively. $\left\langle \psi_{n\mathbf{k}} \mid \psi_{m\mathbf{k+q}}\right\rangle$ denotes the overlap integral between the initial and final electronic wavefunctions $\psi_{n\mathbf{k}}$ and $\psi_{m\mathbf{k+q}}$, which reflects the transition probability between electronic states due to electron-phonon interactions.  
	\begin{figure}[t]
		\centering
		\includegraphics[width=0.45\textwidth]{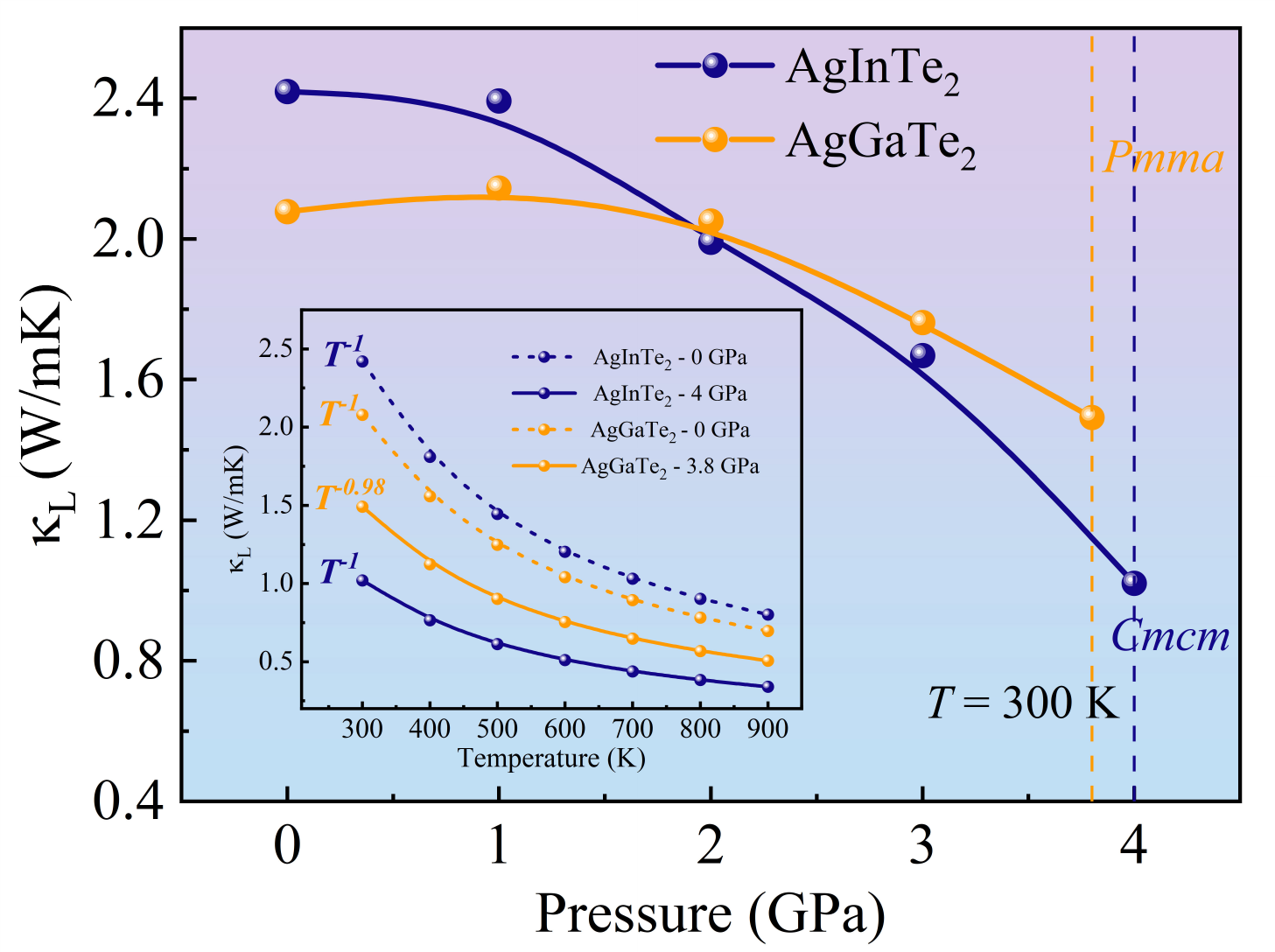}
		\caption{The evolution of lattice thermal conductivity as a function of applied pressure for \ce{AgInTe2} and \ce{AgGaTe2} compounds at 300 K. And the two vertical lines represent the critical phase transition pressures for AgInTe$_2$ and AgGaTe$_2$, respectively.The insert illustrates the temperature-dependent behavior of lattice thermal conductivity for the material under two distinct pressure regimes: ambient pressure and the critical pressure associated with structural phase transition.}
		\label{Fig:2}
	\end{figure}
	
	\begin{figure*}[t]
		\centering
		\includegraphics[width=0.8\textwidth]{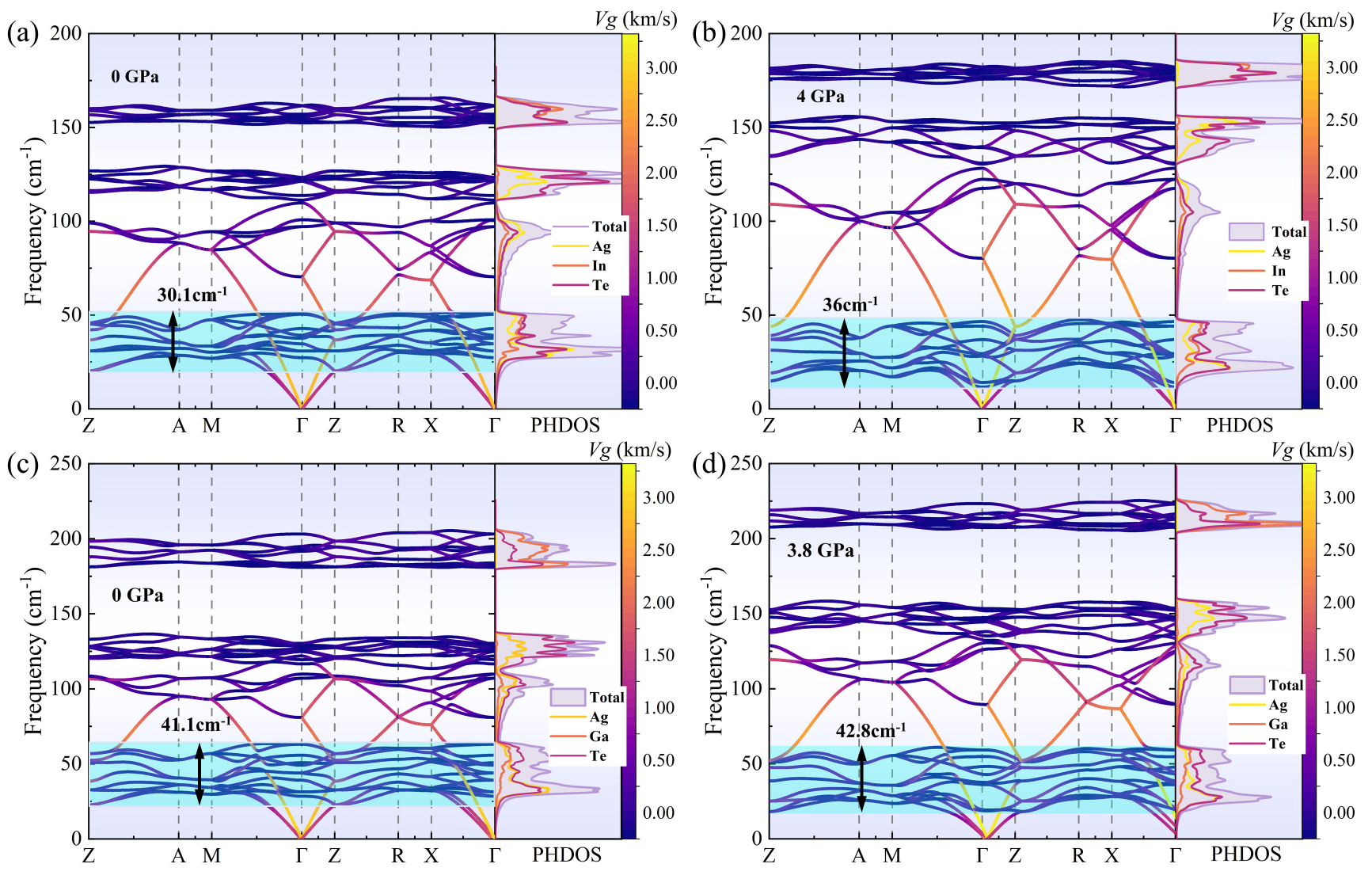}
		\caption{Phonon dispersion curves of (a) $\ce{AgInTe2}$ at 0 GPa, (b) \ce{AgInTe2} at 4 GPa, (c) \ce{AgGaTe2} at 0 GPa, and (d) \ce{AgGaTe2} at 3.8 GPa. The color scale represents the projection of the phonon group velocity for each mode, with maximum group velocities of 3.42 km/s for \ce{AgInTe2} and \ce{AgGaTe2}.}
		\label{Fig:3}
	\end{figure*}
  \section{Resluts and discussion}
  \subsection{Crystal Configuration and Lattice Distortion}  
\ce{AgXTe2} (X=In, Ga) is a tetragonal crystal with a space group of \textit{I}$\bar{4}$2\textit{d}. As illustrated in Fig. \ref{Fig:1}(a), the chalcopyrite structure of A$^I$B$^{III}$X$_{2}^{VI}$ is derived from the binary cubic zincblende structure of ZnSe (\textit{F}$\bar{4}$3\textit{d}). This transformation occurs through the substitution of two Zn atoms with elements from group I (e.g. Cu or Ag) and group III (e.g. Ga, or In). Subsequently, two substituted zincblende unit cells are stacked and extended along the c-direction, resulting in the formation of a tetragonal chalcopyrite structure~\cite{zhang2014high,jaffe1984electronic}.
For \ce{AgXTe2}, it features three distinct types of characteristic tetrahedra: [\ce{TeAg2X2}]$^{6+}$, [\ce{XTe4}]$^{5-}$, and [\ce{AgTe4}]$^{7-}$. These tetrahedra are arranged in a specific manner within the chalcopyrite lattice. The [\ce{XTe4}]$^{5-}$ and [\ce{AgTe4}]$^{7-}$ tetrahedra are linked together to form a 3D framework by sharing corners of the Te atoms~\cite{su2019high,liang2024unveiling}. This structure typically exhibits strong anharmonicity due to the distortion of tetrahedra formed by the linkage of different atoms~\cite{xie2022hidden}.

\par Previous investigations have demonstrated that AgInTe$_2$ and AgGaTe$_2$ maintain structural stability within the pressure ranges of 0-4 GPa and 0-3.8 GPa, respectively~\cite{bovornratanaraks2010high}. Herein, we choose the corresponding phase-transition pressure points to investigate the influence of hydrostatic pressure on the properties of $\ce{AgInTe2}$ and $\ce{AgGaTe2}$. First, we observed that both lattice parameters and bond lengths exhibit varying degrees of reduction under the influence of pressure, as shown in Table~\ref{tbl:1}. (The trend of the lattice constant and volume as a function of pressure is shown in Fig. S1~\cite{supportting}.). Interestingly, the tetrahedral units observed  deviate from ideal ortho-tetrahedral geometry, exhibiting slight distortions, as illustrated in Fig. S2. Generally, the distortions of tetrahedral units are linked to the so-called lattice distortion parameter $\eta=|2-c/a|$, which as an indicator of the deviation degree, as shown in Fig.~\ref{Fig:1}(b). This structural distortion arises from the weak hybridization of \textit{sd\textsuperscript{3}} orbitals of the Ag atom, causing the atoms to deviate from the tetrahedral center~\cite{xie2022hidden}. Consequently, this deviation contributes to the overall distortion of the crystal, leading to strong acoustic and optical phonon scattering. The results demonstrate that the AgGaTe$_2$ system exhibits higher sensitivity to pressure with lattice distortion parameter gradually increases with increasing pressure, indicating that the structure is distorted more under pressure. On the contrary, the lattice distortion parameter of AgInTe$_2$ remains relatively constant under pressure. While structural distortions typically lead to increased phonon scattering, we anticipate a significant reduction in lattice thermal transport under pressure, particularly in the case of \ce{AgGaTe2}.
	\begin{figure*}[t]
		\centering
		\includegraphics[width=1\textwidth]{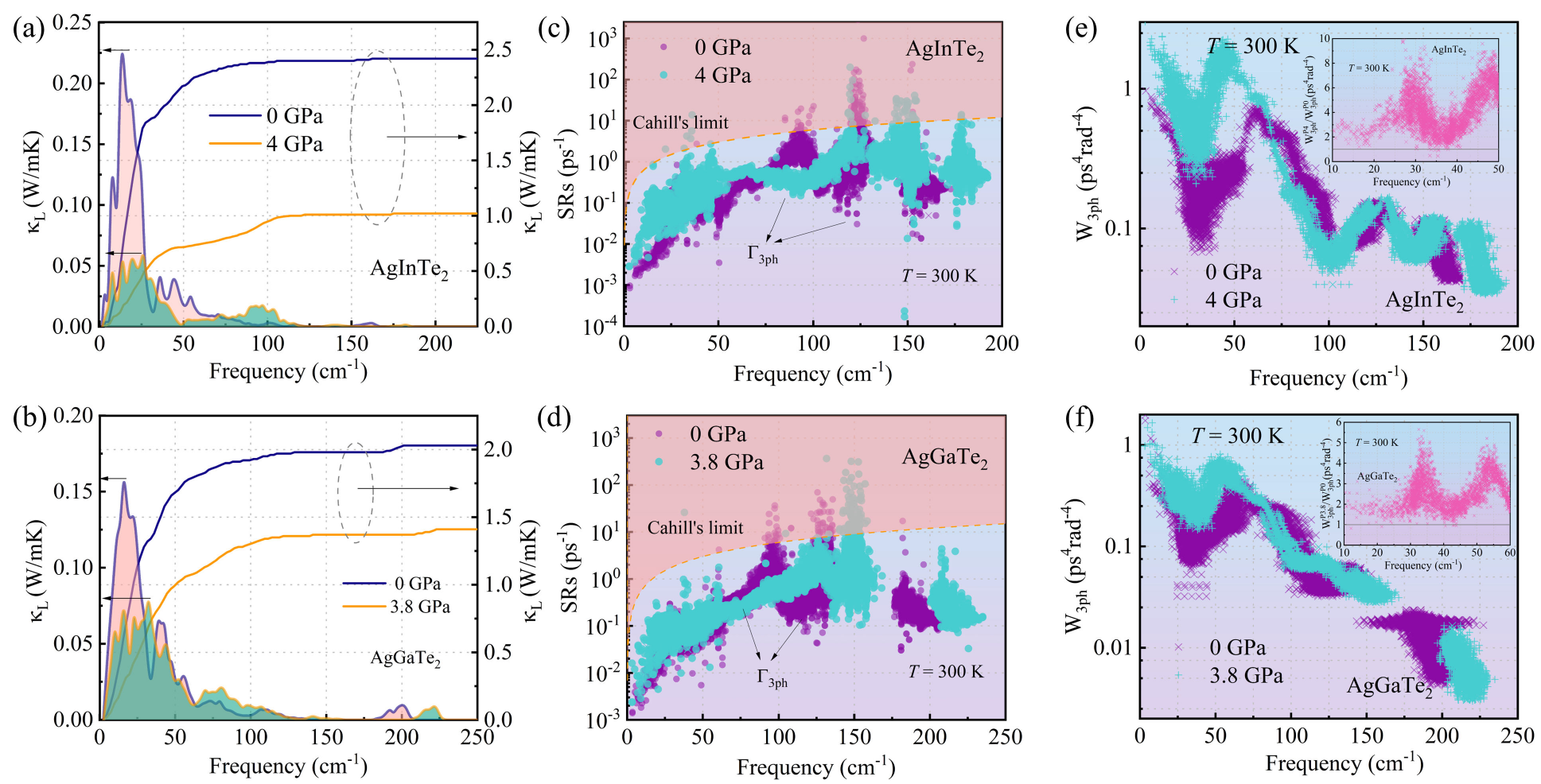}
		\caption{Phonon properties of AgInTe$_2$ and AgGaTe$_2$ were investigated under ambient and high pressure conditions.(a, b) Lattice thermal conductivity spectra and cumulative lattice thermal conductivity with respect to frequency for AgInTe$_2$ and AgGaTe$_2$, revealing the contributions of different phonon modes to heat conduction. The blue line represents 0 GPa, while the yellow line represents the phase transition pressure for each respective material. (c, d) The three-phonon scattering rates for AgInTe$_2$ and AgGaTe$_2$ at 300 K under varying pressures are depicted, with the purple points representing 0 GPa and the blue points corresponding to their respective phase transition pressures. The black dashed line reflects the assumption that the phonon mode scattering rates are twice their frequencies, as predicted by the Cahill-Watson-Pohl model, commonly referred to as the Cahill limit or the amorphous limit.(e, f) Energy- and momentum-conserving three-phonon scattering phase space at 300 K under varying pressure, with the purple points representing 0 GPa and the blue points corresponding to their respective phase transition pressures. The insert figure illustrates the ratio of three-phonon phase space under 4 GPa and 0 GPa conditions, focusing on the low-frequency range.}
		\label{Fig:4}
	\end{figure*}

  \subsection{Lattice Thermal Conductivity}
\par To investigate this correlation, we calculated the lattice thermal conductivity of AgXTe$_2$ (X=In, Ga) under pressure, as illustrated in Fig.~\ref{Fig:2}. The results indicate that the lattice thermal conductivity of AgInTe$_2$ exhibits a monotonically decreasing trend under increasing pressure. Specifically, the $\kappa_{L}$ decreases from 2.42 W/mK at 0 GPa to 1.02 W/mK at 4 GPa. This behavior suggests a continuous reduction in the ability of the lattice to conduct heat as pressure increases, due to enhanced phonon scattering or modifications in the phonon dispersion relations. In contrast, AgGaTe$_2$ shows a non-monotonous trend in its $\kappa_{L}$ under pressure. Initially, the $\kappa_{L}$ increases from 2.08 W/mK at 0 GPa to 2.14 W/mK at 1 GPa, indicating that at lower pressures, there might be a stiffening of the lattice or reduced phonon scattering. However, beyond this point, the thermal conductivity decreases, reaching 1.49 W/mK at 3.8 GPa. This non-linear behavior suggests a more complex interaction between pressure and phonon dynamics, where an initial enhancement in thermal conductivity is followed by a reduction, possibly due to increased anharmonicity or changes in the structural properties at higher pressures. The insert figure illustrates the calculated $\kappa_L$ of AgXTe$_2$ as a function of temperature under different pressures. It can be observed that in the temperature range of 300-800 K, the $\kappa_L$ decreases monotonically with increasing temperature. Additionally, we calculated the temperature dependence of the thermal conductivity under different pressures, and the predicted $\kappa_L$ follows a $T^{-1}$ temperature dependence.

\par Subsequently, we investigated the phonon dispersion curves at different pressures to explain the change, as illustrated in Fig.~\ref{Fig:3} and Fig. S3. 
At ambient pressure, AgXTe$_2$ (X=In, Ga) exhibits flat modes in their low-frequency regions of their phonon dispersion, involving both low-lying optical phonons and acoustic phonons~\cite{tadano2015impact,feldman2000lattice}. This flatness suggests an increased number of phonon modes capable of participating in scattering processes while fulfilling energy and momentum conservation laws, thereby maximizing the phonon-phonon scattering channels~\cite{li2015ultralow}. Notably, the application of pressure induces softening of low-energy phonons and broadens their bandwidth (highlighted by the shaded region in the figure), which encompasses the acoustic phonons and the twelve lowest-lying optical branches. For AgInTe$_2$, this bandwidth expands from 30.1 cm$^{-1}$ to 36 cm$^{-1}$ under pressure. In contrast, the bandwidth expansion in AgGaTe$_2$ is more moderate, increasing from 41.1 cm$^{-1}$ to 42.8 cm$^{-1}$. The manifestation of this change will be demonstrated later, as we currently focus on a simplified phonon relations. 
Assuming the phonon group velocity is given by the derivative of the phonon frequency concerning the wave vector $q$, flat bands correspond to regions of low group velocity across the phonon branches in both AgInTe$_2$ and AgGaTe$_2$. Based on the color mapping of the projected phonon modes, the phonon group velocities in AgInTe$_2$ are slightly lower than those in AgGaTe$_2$ in the low-frequency range below 50 cm$^{-1}$. Upon the application of pressure, both materials exhibit significant softening of their acoustic phonons, accompanied by a modest reduction in group velocity. 
\par In the high-frequency region (above 120 cm$^{-1}$), both materials exhibit a pronounced optical phonon frequency gap, which is attributed to the atomic mass difference between AgInTe$_2$ and AgGaTe$_2$. Upon applying pressures of 4 GPa and 3.8 GPa, both materials exhibit hardening of the high-frequency optical branches. However, the optical phonon frequency gap in AgInTe$_2$ decreases under pressure, whereas it widens in AgGaTe$_2$. This contrasting behavior arises from the faster hardening rate of the high-frequency optical branches compared to the mid-frequency optical branches. In the mid-frequency range (50-120 cm$^{-1}$), we observed a significant hardening of the phonon branches under applied pressure, accompanied by a significant increase in group velocity. This phenomenon arises from the combined influence of the softening of optical branches in the low-frequency regime and the hardening of optical branches in the high-frequency regime, which together account for the substantial increase in group velocity in the mid-frequency region.

\par The lattice thermal conductivity spectra and cumulative thermal conductivity of AgXTe$_2$ under pressure were calculated, as shown in Figs.~\ref{Fig:4}(a-b). At ambient pressure, AgInTe$_2$ exhibited a higher $\kappa_{L}$ of 2.41 W/mK compared to 2.05 W/mK for AgGaTe$_2$. This difference can be attributed to variations in their phonon spectra in the low-frequency region~\cite{yue2024role,yue2024hierarchical}. As illustrated in Fig.~\ref{Fig:4}(a), the $\kappa_{L}$ of AgInTe$_2$ is predominantly determined by phonons in the frequency range of 0-30 cm$^{-1}$. When subjected to pressure, a notable alteration in the thermal conductivity spectrum is observed. At 4 GPa, the frequency range contributing to $\kappa_{L}$ expands to 0-50 cm$^{-1}$. This expansion indicates an increase in the phonon scattering rate and available phase space within this extended range. Such findings are consistent with the observed softening and broadening of low-energy phonons in the phonon spectra. In contrast, as illustrated in Fig.~\ref{Fig:4}(b), the thermal conductivity spectrum of AgGaTe$_2$ at 3.8 GPa showed relatively minor changes in the low-frequency region. 
We observed a slight enhancement in the thermal conductivity contribution within the mid-frequency range (80-120 cm$^{-1}$). This observation can be attributed to the augmented phonon group velocity in the corresponding mid-frequency region of the phonon dispersion curves (Fig. 3), which aligns with previous analyses. And the AgInTe$_2$ is same obsertion applies as well. Furthermore, additional analysis indicates that the pressure dependence of $\kappa_{L}$ in the two materials exhibits significant discrepancies. AgInTe$_2$ exhibits a more pronounced decrease in $\kappa_{L}$ under pressure, dropping by 57\% at 4 GPa, while the $\kappa_{L}$ of AgGaTe$_2$ decreases more moderately, by 28\% at 3.8 GPa. This disparity can be attributed to variations in phonon scattering rates and phase space under pressure for the two materials.
\begin{figure*}[t]
		\centering
		\includegraphics[width=1\textwidth]{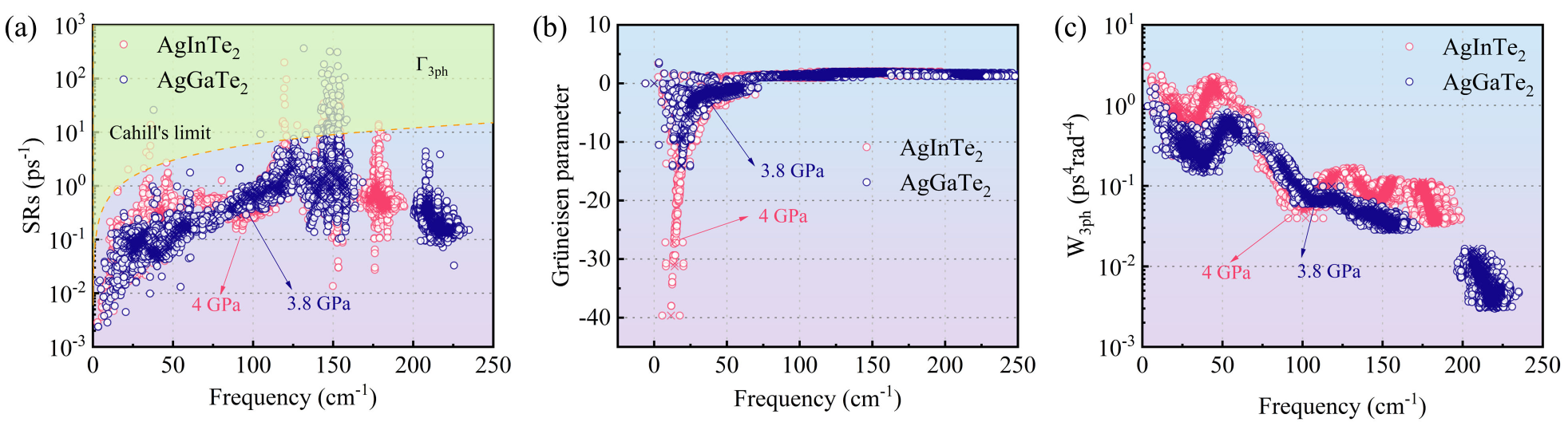}
		\caption{A comparative analysis of the phonon scattering properties of AgInTe$_2$ (red circles) and AgGaTe$_2$ (blue circles) at their respective phase transition pressures. (a) Three-phonon scattering rates; the orange dashed line reflects the assumption that the phonon mode scattering rates are twice their frequencies, commonly referred to as the Cahill limit.(b) Grüneisen parameters. (c) Available phase space for three-phonon scattering processes. These properties provide insights into the anharmonic lattice dynamics of each material.}
		\label{Fig.5}
\end{figure*}
        \begin{table*}[t]
		\centering
		\caption{The transferred charge and shared charges ($\left|e\right|$) of \ce{AgInTe2} and \ce{AgGaTe2} at 0 and 4 (3.8) GPa calculated using chargmol.}
		\label{Table:2}
		\begin{tabular}{ccccccc}
			\hline\hline
			& \multirow{2}{*}{Pressure (GPa)} & \multicolumn{3}{c}{Transferred charge ($\left|e\right|$)} & \multicolumn{2}{c}{Shared charges ($\left|e\right|$)} \\
			\cmidrule(lr){3-5} \cmidrule(lr){6-7}
			& & Ag & In or Ga & Te & Ag-Te & In/Ga-Te \\
			\midrule
			\ce{AgInTe2} & 0   & 0.191 & 0.542 & -0.367 & 0.914 & 0.970 \\
			& 4   & 0.196 & 0.552 & -0.374 & 1.018 & 1.041 \\
			\ce{AgGaTe2} & 0   & 0.165 & 0.420 & -0.293 & 0.910 & 0.992 \\
			& 3.8 & 0.172 & 0.426 & -0.299 & 0.995 & 1.053 \\
			\hline\hline
		\end{tabular}
	\end{table*}

\par To elucidate the observed discrepancies, the three-phonon scattering rates and scattering phase spaces were calculated for both materials under varying pressure conditions, as illustrated in Figs.~\ref{Fig:4}(c-f). Figs.~\ref{Fig:4}(c-d) clearly demonstrate that upon compression, both materials exhibit an increased phonon scattering rate at the crossover between the acoustic and optical branches, albeit to differing degrees. This increase would contribute to the reduction in lattice thermal conductivity. Notably, this frequency range lies at the boundary between acoustic and low-frequency optical phonons, suggesting that the coupling between acoustic and low-frequency optical phonons is enhanced under pressure. The prominent contribution of Ag and Te atomic phonon density of states at the acoustic-optical phonon crossover in the phonon dispersion curves substantiates that Ag and Te atom-induced low-lying optical phonon branches are responsible for the strong coupling between acoustic and optical phonons, as shown in Fig.~\ref{Fig:3}. The coupling of these modes contributes to an expanded scattering phase space and leads to additional scattering events.~\cite{yue2024high} The changes in phase space indicate that modes with larger phase space facilitate a greater number of scattering events, resulting in higher scattering rates. At a pressure of 4 GPa, AgInTe$_2$ exhibits a significant expansion of its scattering phase space within the low-frequency region of 0-75 cm$^{-1}$, as shown in Fig.~\ref{Fig:4}(e). This expansion reflects a substantial increase in the number of three-phonon interaction channels satisfying energy and momentum conservation within this frequency range, a prerequisite for enhanced three-phonon scattering processes. While AgGaTe$_2$ also exhibits a moderate increase in scattering phase space at a pressure of 3.8 GPa, the magnitude of this change is considerably smaller compared to that observed in AgInTe$_2$. 
	\begin{figure*}[t]
		\centering
		\includegraphics[width=0.8\textwidth]{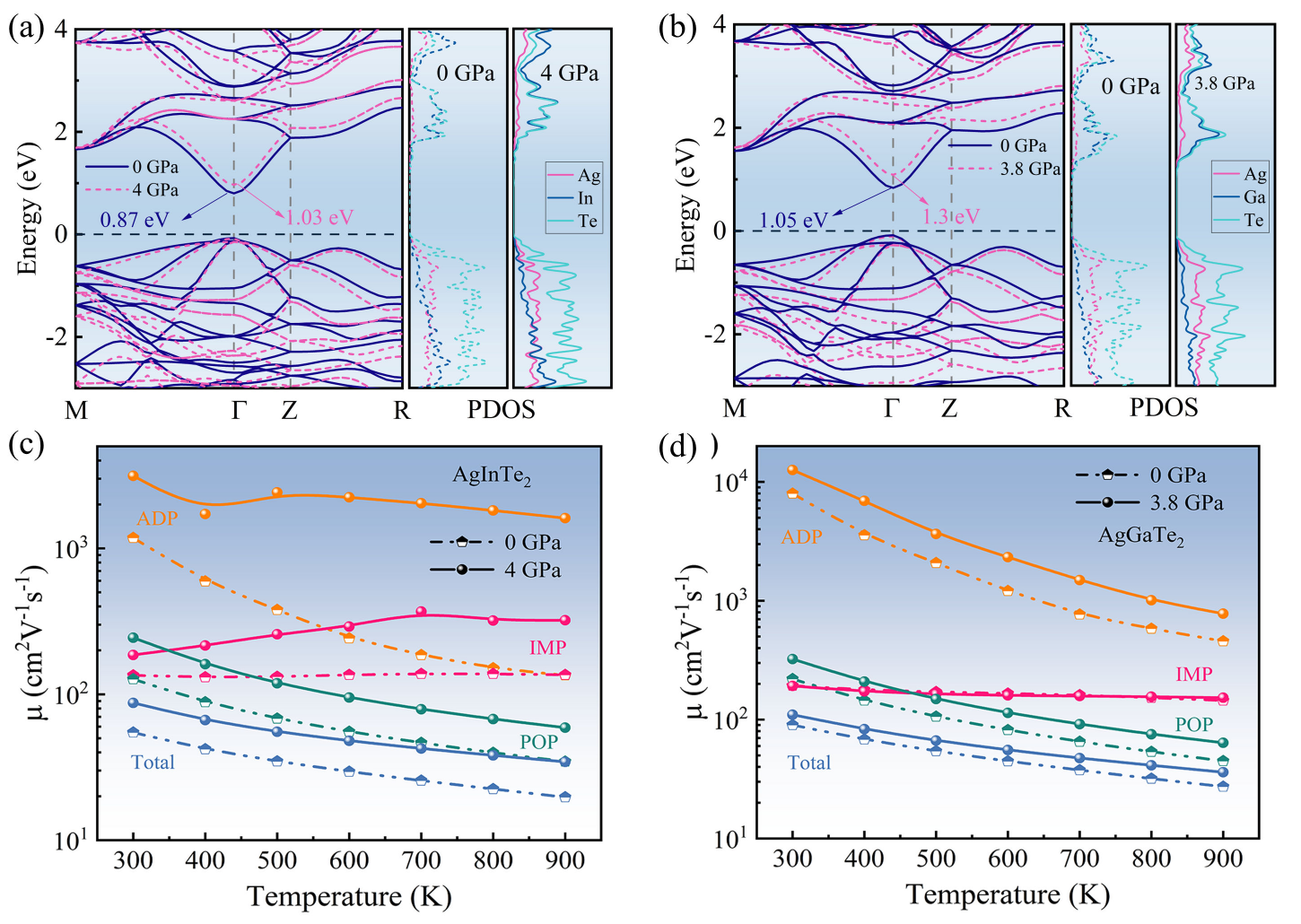}
		\caption{Pressure-Dependent Electronic Structure and Carrier Mobility in AgInTe$_2$ and AgGaTe$_2$. (a)-(b) Calculated band structures and densities of states at atmospheric pressure (dashed lines) and the phase transition pressure (solid lines). (c)-(d) Calculated carrier mobility, including contributions from different scattering mechanisms and the total mobility, at 0 GPa (dashed lines) and the phase transition pressure (solid lines).}
		\label{Fig.6}
	\end{figure*}
	
\subsection{Pressure-induced Metavalent Bonding and Anharmonicity}
\par The degree of lattice distortion within a crystal structure plays a significant role in phonon transport. Generally, a higher degree of lattice distortion corresponds to lower $\kappa_{L}$. This relationship can be summarized as: increased structural complexity leads to reduced $\kappa_{L}$. Under ambient pressure conditions, both AgGaTe$_2$ and AgInTe$_2$ appear to follow this trend, with the more distorted AgGaTe$_2$ exhibiting lower $\kappa_{L}$. However, this correlation does not hold true under high pressure. Despite exhibiting a larger lattice distortion parameter than AgInTe$_2$ at high pressure, AgGaTe$_2$ surprisingly demonstrates higher $\kappa_{L}$. This deviation suggests that under high-pressure conditions, factors beyond lattice distortion, such as changes in phonon scattering mechanisms or electronic contributions to thermal conductivity, may become more dominant in influencing $\kappa_{L}$.
\par To explore the underlying mechanisms responsible for the observed significant discrepancies in the lattice thermal conductivity of \ce{AgInTe2} and \ce{AgGaTe2} under pressure, a comparative analysis of their transport properties under pressure was conducted, as illustrated in Fig.~\ref{Fig.5}.  As shown in Fig.~\ref{Fig.5}(a), the three-phonon scattering rates of AgInTe$_{2}$ and AgGaTe$_{2}$ at room temperature of 4 GPa and 3.8 GPa, respectively. We find that the scattering rate of \ce{AgInTe2} is much larger than that of \ce{AgGaTe2} in the low-frequency region, which explains why the change in thermal conductivity of \ce{AgInTe2} is more sensitive at high pressures. In order to explore which parameter has a greater effect on the scattering rate, we discuss below the effect of the three-phonon scattering phase space and the Grüneisen parameter on the scattering rate. 
The Grüneisen parameter, $\gamma_{\lambda}$, is expressed as:
	\begin{equation}
		\begin{aligned}
			\gamma_{\lambda} &= -\frac{1}{6\omega_{\lambda}^{2}} \sum_{k} \sum_{l' k'} \sum_{l'' k''} \sum_{\alpha \beta \gamma} \\
			& \times \left[ \Phi_{\alpha \beta \gamma}(0k, l' k', l'' k'') 
			\frac{e_{\alpha k}^{\lambda *} e_{\beta k'}^{\lambda'}}{\sqrt{M_{k} M_{k'}}} e^{iqR_{l'}} r_{l'' k''}^{\gamma} \right]
		\end{aligned}
	\end{equation}
where $\Phi_{\alpha \beta \gamma}(0k, l' k', l'' k'')$ is third-order interatomic force constant (IFC) tensor, $e_{\alpha k}^{\lambda }$ is polarization vector of the phonon mode $\lambda$ at atomic site $k$ in the $\alpha$ direction, $e_{\beta k'}^{\lambda'}$ is polarization vector of another phonon mode $\lambda'$ at atomic site $k'$ in the $\beta$ direction, $e^{iqR_{l'}}$ is a phase factor related to the position of the atom at site $l'$, $r_{l'' k''}^{\gamma}$ is displacement vector of the atom at site $l''$ in the $\gamma$ direction. It is worth noting that the Grüneisen parameter of \ce{AgInTe2} is very much larger than that of \ce{AgGaTe2}, about more than three times that of \ce{AgGaTe2}, which implies a greater anharmonicity of AgInTe$_2$. 
       \begin{figure*}[t]
		\centering
		\includegraphics[width=0.8\textwidth]{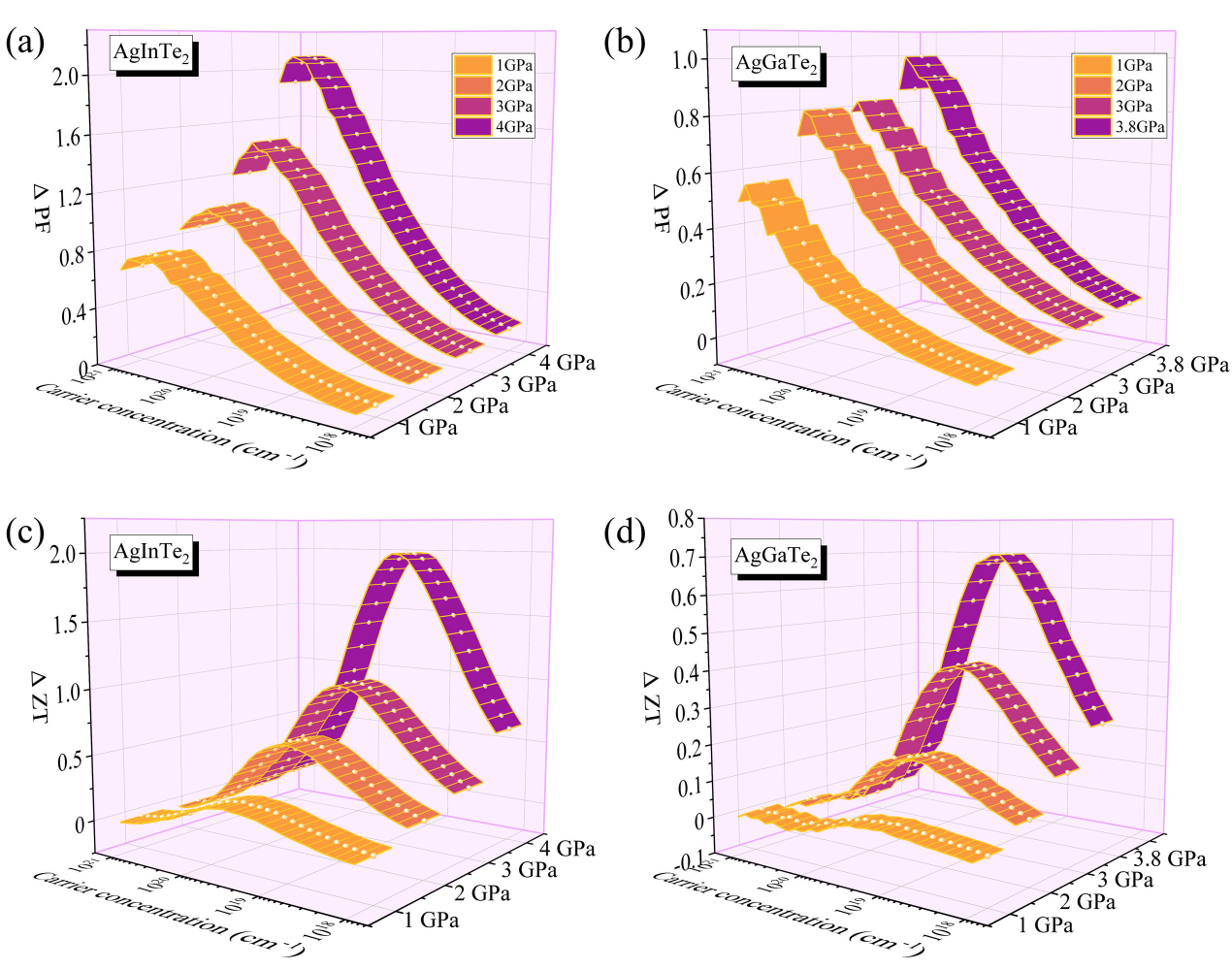}
		\caption{The calculated \textit{$\Delta$PF} of (a) AgInTe$_2$ and (b) AgGaTe$_2$ as a function of carrier concentration and pressure at 900 K. The \textit{$\Delta$ZT} of (c) AgInTe$_2$ and (d) AgGaTe$_2$ as a function of carrier concentration and pressure at 900 K. Where $\Delta$$PF$ = $PF_{P}$ - $PF_{0}$, $\Delta$$ZT$ = $ZT_{P}$ - $ZT_{0}$, the change in color in the graph indicates the magnitude of the \textit{$\Delta$$PF$} and \textit{$\Delta$$ZT$} changes.}
		\label{Fig.7}
	\end{figure*}
\par The phase space (W$^{\pm}_{\lambda}$) for three-phonon scattering is defined as the sum of the frequency factors associated with the expression ( $\Gamma^{\pm}_{\lambda \lambda \lambda}$ ), given by:
       \begin{equation}
	\begin{aligned}
		W_{\lambda}^{\pm}  &= \frac{1}{2N} \sum_{\lambda' \lambda''} 
		\left\{
		\begin{matrix}
			2(f_{\lambda'} - f_{\lambda''}) \\
			f_{\lambda'} + f_{\lambda''} + 1 
		\end{matrix}
		\right\} \\
        &\quad \times \frac{\delta (\omega_{\lambda} \pm \omega_{\lambda'} - \omega_{\lambda''})}{\omega_{\lambda} \omega_{\lambda'} \omega_{\lambda''}}
	\end{aligned}
       \end{equation}
	Where \( f_{\lambda'} \) and \( f_{\lambda''} \) are Fermi-Dirac distribution functions, \( \delta (\omega_{\lambda} \pm \omega_{\lambda'} - \omega_{\lambda''}) \) is the Dirac delta function, the denominator \( \omega_{\lambda} \omega_{\lambda'} \omega_{\lambda''} \) is the product of the frequencies, typically used for normalization or representing the reciprocal of energy units. The phonon scattering phase space at high pressure shows that the data obtained for \ce{AgInTe2}, is much larger than that of \ce{AgGaTe2}, indicating that the phonon dispersion curves of \ce{AgInTe2} at pressure better satisfy the energy conservation and momentum conservation.
 
  To further investigate the bonding characteristics, we computed the shared and transform charge for both compounds, as depicted in Table~\ref{Table:2}. Analysis indicates a stronger Ag-Te bond in AgInTe$_2$ at ambient pressure than in AgGaTe$_2$. Conversely, the Ga-Te bond exhibits a greater degree of charge sharing than the In-Te bond, suggesting a slightly stronger interaction in the former case. Notably, the In compound exhibits a larger charge transfer than the Ga compound, which can be attributed to the higher electronegativity of Ga. In addition, both compounds show an increasing trend in charge transfer with increasing pressure. Metavalent bonding, distinct from covalent or metallic bonding, characterizes these compounds. A defining feature of metavalent bonds is their high degree of anharmonicity, which exhibits a positive correlation with the extent of charge transfer within the compound. The deviation from an ideal tetrahedral structure underpins the formation of metavalent bonds. Consequently, \ce{AgInTe2} exhibits stronger anharmonicity than \ce{AgGaTe2} under both ambient and high-pressure conditions. However, it is noteworthy that despite its weaker anharmonicity, $\ce{AgGaTe2}$ displays lower lattice thermal conductivity than $\ce{AgInTe2}$ under ambient pressure. This phenomenon can be attributed to the more complex structural characteristics of $\ce{AgGaTe2}$ under standard conditions.
	 
  \subsection{Electrical Transport and Electron-phonon Scattering}
\par Next we explore the effect of pressure on the electronic structure of \ce{AgInTe2} and \ce{AgGaTe2}. Figs.~\ref{Fig.6}(a-b) displays the electronic band structure and the density of states under hydrostatic pressure. The computed bandgap values at atmospheric pressure are 0.87 eV for \ce{AgInTe2} and 1.03 eV for \ce{AgGaTe2}, respectively. Due to the quantum size effect~\cite{ekimov1985quantum}, the application of pressure causes a slight increase in the band gap, which increases to 1.05 eV and 1.3 eV at 4 GPa and 3.8 GPa, respectively. Analogous behavior is exhibited in other chalcopyrite materials, such as \ce{CuInTe2}, where the band gap undergoes a notable expansion from 1.03 eV to 1.44 eV under an applied pressure of 7.6 GPa~\cite{yue2023pressure}. The pressure coefficient of the bandgap, denoted as $\Delta E_{g}/\Delta P$, quantifies the responsiveness of the bandgap to variations in applied pressure. For AgInTe$_2$ and AgGaTe$_2$, the respective pressure coefficients for bandgap widening are 45 meV/GPa and 71 meV/GPa. The notably higher value for AgGaTe$_2$ suggests a greater sensitivity of its bandgap to pressure-induced changes compared to AgInTe$_2$. This difference in pressure sensitivity may arise from variations in the compressibility and bonding characteristics between the two materials, leading to distinct responses to applied pressure in terms of their electronic properties. In addition, the valence band becomes more dispersive under pressure, which will greatly improve the increase in $\sigma$. In addition, from the density of states in the electronic band structures, it is evident that the valence band maximum in both \ce{AgInTe2} and \ce{AgGaTe2} is primarily contributed by Te and Ag atoms, with a negligible contribution from In atoms. This contribution remains largely unchanged under the influence of applied pressure.
	
\par Furthermore, pressure not only preserves but also improves the electronic transport properties. To illustrate this phenomenon, the hole mobility was calculated by incorporating ADP, POP, and IMP scattering mechanisms to accurately model electrical transport characteristics. Figs.~\ref{Fig.6}(c-d) illustrates the total and partial hole mobilities limited by the respective scattering mechanisms. In AgInTe$_2$, the magnitudes of the three scattering mechanisms are comparable, resulting in a significant influence on the total hole mobility, as illustrated in Fig.~\ref{Fig.6}(c). Under pressure, ADP scattering increases substantially, followed by enhancements in IMP and POP scattering, resulting in a marked improvement in total hole mobility. For AgGaTe$_2$, although the improvement in IMP scattering due to pressure is minimal, the substantial enhancements in ADP and POP significantly contribute to the overall increase in hole mobility. The significant enhancement in hole mobility is attributed to the combined effect of reduced single-band effective mass and deformation potential constant, stemming from pressure-induced tetrahedral distortion~\cite{cao2018dominant}. In the chalcopyrite systems under investigation, POP scattering emerges as the dominant mechanism controlling hole mobility, consistent with recent reports on polar systems such as PbTe~\cite{zhou2018half-Heuslers}, SnSe~\cite{ma2018SnSe}, ReCuZnP$_2$~\cite{RECuZnP22021experimental}, and several half- and full-Heusler compounds~\cite{guo2024novel}. Additionally, we calculated $\textit{$\kappa_e$}$, $\textit{S}$, $\textit{$\kappa_L$}$, and $\textit{PF}$, as presented in Fig. S4 of the supplementary material.
	
 \begin{figure}[t]
		\centering
		\includegraphics[width=0.47\textwidth]{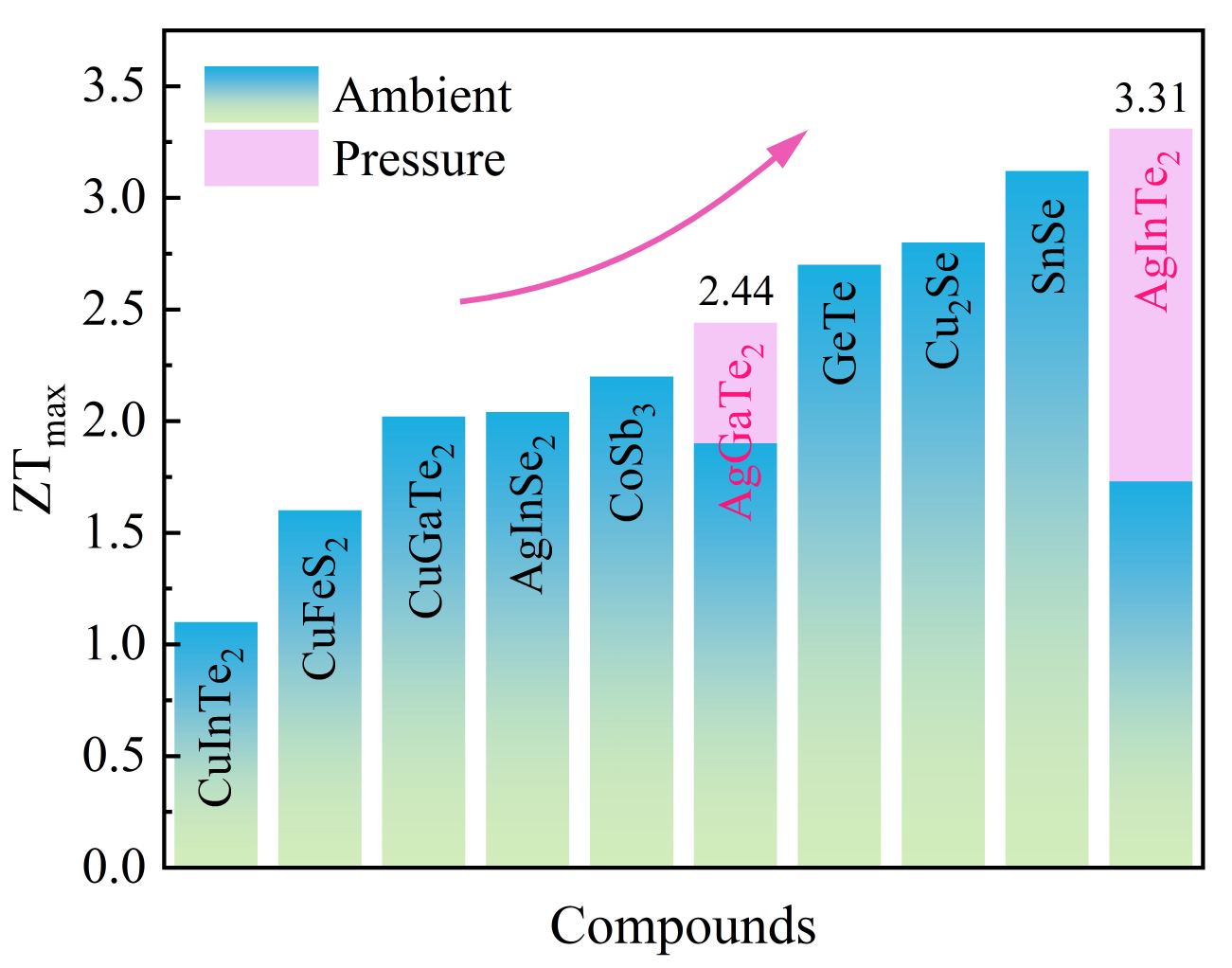}
		\caption{Maximum ZT values of AgXTe$_2$ (X=In, Ga) [This work], CuInTe$_2$~\cite{yue2023pressure}, CuFeS$_2$~\cite{li2019high}, CuGaTe$_2$~\cite{chen2019computational}, SnSe, Cu$_2$Se, GeTe, and CoSb$_3$~\cite{tang2024high} under ambient and applied pressure.  Ambient pressure results are indicated by the blue-green gradient, while pressure-tuned results are represented by the purple section.}
		\label{Fig.8}
	\end{figure}
 \subsection{Power Factor and Figures of Merit}
\par Subsequently, we conducted an extensive analysis of the pressure-dependent $PF$ performance to elucidate this relationship quantitatively. Figs.~\ref{Fig.7}(a-b) showcase heat maps that vividly illustrate the intricate variations in carrier concentration and power factor change (\textit{$\Delta$PF = PF$_P$ - PF$_0$}) as a function of applied pressure at 900 K. Our findings reveal a striking contrast between AgInTe$_2$ and AgGaTe$_2$. The former exhibits a remarkably larger PF enhancement, with a maximum increase of approximately 2.1 mW m$^{-1}$ K$^{-2}$ at a carrier concentration of 6.21 $\times$ 10$^{20}$ cm$^{-3}$. This substantial improvement underscores the potential of AgInTe$_2$ as a highly responsive material to pressure-induced optimization. Conversely, AgGaTe$_2$, while demonstrating a more modest enhancement, still achieves a noteworthy optimal increase of about 1 mW m$^{-1}$ K$^{-2}$ at 7.88 $\times$ 10$^{20}$ cm$^{-3}$ and 3.8 GPa. Intriguingly, both compounds exhibit maximum optimization near their respective phase transition pressures and in heavily doped regions. This convergence of optimal conditions suggests a principle in pressure-tuned thermoelectrics: the synergistic effect of structural changes and high carrier concentrations can lead to significant performance enhancements. Such a finding opens new avenues for designing high-performance thermoelectric materials tailored for extreme conditions.
	
\par To provide a holistic view of the thermoelectric potential, we extended our analysis to the thermoelectric \textit{ZT}, a crucial parameter for assessing overall thermoelectric efficiency. Figs.~\ref{Fig.7}(c-d) present our predictions for \textit{$\Delta$ZT} (\textit{$\Delta$ZT = ZT$_P$ - ZT$_0$}) at 900 K as a function of hole concentration, offering insights into the optimal doping levels under pressure. Furthermore, the dependence of the thermoelectric $ZT$ of AgXTe$_2$ on carrier concentration under various pressures was calculated, as illustrated in Supplementary Fig. S5. The results for AgInTe$_2$ are particularly striking. We observe a remarkable \textit{ZT} enhancement of 1.98 at a hole concentration of 2.21 $\times$ 10$^{19}$ cm$^{-3}$ and 900 K. Even more impressive is the peak \textit{ZT} value of 3.31, achieved under an applied pressure of 4 GPa and a concentration of 3.56 $\times$ 10$^{19}$ cm$^{-3}$. This surprising enhancement showcases the transformative potential of pressure tuning in thermoelectric materials. While the \textit{ZT} enhancement in AgGaTe$_2$ is less dramatic, it is not insignificant. The inherently excellent thermoelectric properties of the compound at ambient pressure offer a robust foundation. Coupled with pressure modulation, this results in a maximum $ZT$ value of 2.44 at 900 K and a hole concentration of $4.52 \times 10^{19}$ cm$^{-3}$. In Fig.~\ref{Fig.8}, we compare the $ZT$ values of AgXTe$_2$ (X=In, Ga) with those of conventional thermoelectric materials~\cite{yue2023pressure,li2019high,chen2019computational,tang2024high}. The data indicate that pressure-tuned AgInTe$_2$ achieves competitive $ZT$ values, reinforcing its potential as a high-performance thermoelectric material. Our result underscores the importance of considering both intrinsic material properties and external tuning parameters in the quest for high-performance thermoelectrics. These findings not only clarify the pressure-dependent behavior of AgXTe$_2$ (X=In, Ga) compounds but also contribute to the broader understanding of structure-property relationships in thermoelectric materials.

\section{Conclusions}	
\par In summary, the thermal transport and electrical transport properties of Ag-based chalcopyrite materials were studied using the first-principles calculation combined with the Boltzmann transport equation. The results reveal that AgInTe$_2$ exhibits superior lattice rigidity and consequently undergoes less lattice distortion under pressure compared to AgGaTe$_2$, which possesses weaker lattice rigidity. This phenomenon can be attributed to the stronger bonding strength of Ag-Te in the AgInTe$_2$ lattice compared to that in AgGaTe$_2$, resulting in a more resilient structure that is less susceptible to deformation under applied pressure. Moreover, the pressure-induced softening of low-frequency phonons significantly increases the low-frequency three-phonon scattering channels, greatly enhancing the coupling between acoustic and optical phonons, leading to a sharp decrease in lattice thermal conductivity. Interestingly, the positive correlation between lattice distortion and lattice thermal conductivity observed at ambient pressure does not appear to hold under high pressure. Despite maintaining structural symmetry under high pressure, AgInTe$_2$ exhibits lower lattice thermal conductivity than AgGaTe$_2$, which has a higher degree of lattice distortion, due to the enhanced anharmonicity induced by high pressure. Under the influence of hydrostatic pressure, the $ZT$ values of AgInTe$_2$ and AgGaTe$_2$ exhibited remarkable enhancements. Surprisingly, the maximum $ZT$ value of AgInTe$_2$ increased by approximately 90\%\, while that of AgGaTe$_2$ improved by around 27\%. These significant improvements in the thermoelectric performance under hydrostatic pressure conditions underscore the potential of pressure-induced optimization in chalcopyrite-type thermoelectric materials. The significant enhancements observed under pressure highlight the potential for developing next-generation thermoelectric devices capable of operating efficiently under extreme conditions, paving the way for novel applications in energy harvesting and thermal management.

\section{Acknowledgments}
The authors thank the National Natural Science Foundation of China (No. 52072188), the Program for Science and Technology Innovation Team in Zhejiang (No. 2021R01004), and acknowledge the institute of high-pressure physics of Ningbo University for its computational resources.

	\nocite{*}
	
	\bibliography{apssamp}% Produces the bibliography via BibTeX.
	
\end{document}